\documentclass[doublecol]{epl2}
\usepackage[T1]{fontenc}

\usepackage{amssymb}
\usepackage{amsmath}
\renewcommand{\b}{\textbf}

\title{Electric dipoles vs. magnetic dipoles - for two molecules in a harmonic trap}
\shorttitle{Electric dipoles vs. magnetic dipoles} 

\author{Wojciech G\'orecki\inst{1} \and Kazimierz Rz\k{a}\.zewski\inst{1}}
\shortauthor{Wojciech G\'orecki \etal}

\institute{                    
  \inst{1} Center for Theoretical Physics, Polish Academy of Sciences - Al. Lotnik\'ow 32/46, 02-668 Warsaw, Poland\\
}

\abstract{We study energy levels of two heteronuclear molecules moving in a spherically symmetric harmonic trap. A role of electric dipole interactions is compared and contrasted with our earlier results \cite{my} for two magnetic dipolar atoms. We stress importance of a rotational energy with its value which is very high compared to the energy of dipolar interaction. We show that dipolar forces do not play a significant role in the ground state of the system under typical experimental conditions. However, there exist excited states that exhibit anticrossings similar to the ones observed for magnetic dipoles.}

\begin{document}

\maketitle

\section{Introduction}
Since a successful condensation of chromium \cite{chrom2,chrom1}, magnetic dipoles interaction became an important topic in physics of quantum gases. The reason for this is its significant influence on the behaviour of Bose-Einstein condensate, which can be also observed for other atoms with large magnetic moment, like erbium \cite{erbium} or dysprosium \cite{dysp1,dysp2}. Beside theoretical works on large condensates \cite{EdH1,EdH2,chredh,rubid1,rubid2,rubid3}, there have appeared papers on a few (even two \cite{2atomy1,2atomy2,my}) interacting particles. Moreover, due to a 
progress in controlling single atoms in optical lattices \cite{opty1,opty2}, we hope that in the near future it will be possible to check all these predictions in experiments.\par
Simultaneously, we observe a significant progress in production of ultracold samples of dipolar heteronuclear molecule. The gas may be cooled to $mK$ regime in laboratory \cite{ck1,ck2,ck3,ck4}; recently even samples in the absolute rovibrational ground state are created \cite{cg1,cg2,cg3,cg4,cg5}. As in the classical physics the electric dipole interaction is much stronger than the magnetic dipole interaction - i.e. for the electric dipole of value $d\approx1D$ (which is a typical order of magnitude for heteronuclear molecules like HCl, HF etc.) we have $\dfrac{d_1d_2}{4\pi\epsilon_0r^3}\approx10^4\dfrac{\mu_0\mu_B}{4\pi r^3}$ - it seems to be very promising object to analyze, because all interesting effects should be visible very clearly for such strong interactions.\par
Nonetheless, there are relatively few theoretical papers on electric dipoles. Even when the electric dipole interaction is considered, there are some restrictions: the gas is strongly polarized \cite{elec} or very specific traps are considered \cite{2d}; there are also papers on building a quantum computer by using heteronuclear molecules \cite{comp}. Simulating of some condensed matter model with tunable parameters by using diatomic polar molecules in optical lattice is discussed as well \cite{a1,a2}. In all these cases the number of  degrees of freedom which dipolar molecule naturally possess is diametrically restricted. On the contrary, in this Letter we want to focus our attention on the case without an external electric field and make no assumptions on molecules' position or orientation is space. Especially we are interested in analyzing the effect which was described in Ref. \cite{my}. It turns out, that for totally spherically symmetric system of two magnetic dipolar atoms in a harmonic trap it is possible to generate a non-zero relative orbital angular momentum by changing the strength of dipole-dipole interaction (which may be effectively done by manipulating a trap frequency). This effect results from the principle of total angular momentum conservation (while the dipolar interaction couples spin with the orbital angular momentum). In some sense it is an analogue of the Einstein de Haas effect \cite{edh}, where a rotation of the system is generated by changing its magnetisation by using an external magnetic field. However, there is a significant difference - for the EdH effect the fact, that magnetization is a pseudovector is crucial, so it is obvious that it can not have analogue for electric dipoles (as electric field or electric dipoles are just vectors). On the other hand, in the case of Ref. \cite{my} no direction is specified (only the  $\langle L^2\rangle$ is generated, but all $\langle L_x\rangle$, $\langle L_y\rangle$, $\langle L_z\rangle$ remain equal to zero), so there is a chance to find some analogues effect.\par
In this Letter we start from fundamentals in analyzing the nature of the electric dipoles, which is diametrically different than the magnetic dipoles. Nonetheless, it turns out, that analogous effects connected with the generation of a non zero $\langle L^2\rangle$ occurs, even if this will be rather hard to observe in the experiment.

\section{Rigid rotators approximation}In contrast to magnetic dipole, the electric dipole is not an elementary object in quantum mechanics. To analyze its properties we need to start with the whole Hamiltonian of a dipolar molecule and then make proper approximations. As the case without external electric field is under consideration, we focus on constant electric dipoles and we neglect the induced electric dipole moment. For simplicity, we will consider diatomic molecules. By using the Born-Oppenheimer approximation, we have three distinguishable components of the spectrum - rotational states, vibrational states and electronic states. Typical values of the energy gap between the lowest states are respectively $0.1-10cm^{-1}$ (rotational), $10^2-10^3cm^{-1}$ (vibrational) and $10^6cm^{-1}$ (electronic). For comparison, for the harmonic oscillator with a frequency $\omega=2.8\cdot2\pi kHz$ the value of the energy gap $\hbar\omega\approx10^{-7}cm^{-1}$.\par
To decide, which parts of the energy must be included, we need to consider the form of the dipole operator $\b{d}_i$:
\begin{equation}
\b{d}_i=d_i\cdot\b{e}_i=d_i\cdot\big(sin(\theta_i)cos(\varphi_i),sin(\theta_i)sin(\varphi_i),cos(\theta_i)\big)
\end{equation}
where $i$ is the index denoting $i^{th}$ molecule.
One can see, that for states with the well defined orbital angular momentum $|l_i,m_{i}\rangle$ we have:
\begin{equation}
\langle l_i',m_{i}'|\b{d}_i|l_i,m_{i}\rangle\neq 0 \Rightarrow |l_i'-l_i|=1
\end{equation}
Therefore considering rotational states is crucial in the problem of electric dipoles interactions; two others (vibrational and electronic) are not so important and, as they are much higher in energy, they may be neglected, since we may assume that our molecules remain in their respective ground states. Thus, we will treat the molecules as rigid rotators:
\begin{equation}
H_{roti}=\frac{\hbar^2}{2I}\b{L}^2_i
\end{equation}
where $\b{L}_i$ is the dimensionless orbital angular momentum operator. This equation is not exact, as in reality the value of molecule's moment of interia $I$ is not constant (it slowly grows with $l$ due to the centrifugal force) but for our considerations this approximation is good enough. The energy of dipole interaction between the molecules is given as:
\begin{equation}
H_{dd}=\frac{d_1d_2}{4\pi\epsilon_0|\b{r}_1-\b{r}_2|^3}\big(\b{e}_1\cdot\b{e}_2-3(\b{e}_1\cdot\b{n})(\b{e}_2\cdot\b{n})\big)
\end{equation}
where
\begin{equation}
\b{n}=\frac{\b{r}_1-\b{r}_2}{|\b{r}_1-\b{r}_2|}=\big(sin(\theta)cos(\varphi),sin(\theta)sin(\varphi),cos(\theta)\big)
\end{equation}
Finally, a simplified Hamiltonian of two identical electric dipoles in a harmonic trap can be written down as (in dimensionless oscillator units):
\begin{multline}
H=-\frac{1}{2}\Delta_1-\frac{1}{2}\Delta_2+\frac{1}{2}r_1^2+\frac{1}{2}r_2^2+ B \b{L}_1^2+ B \b{L}_2^2+\\\frac{\tilde{g}_{dd}}{|\b{r}_1-\b{r}_2|^3}\big(\b{e}_1\cdot\b{e}_2-3(\b{e}_1\cdot\b{n})(\b{e}_2\cdot\b{n})\big)+V_{SR}(\b{r}_1,\b{r}_2)
\end{multline}
where $ B =\frac{\hbar}{2I\omega}$, $\tilde{g}_{dd}=\frac{d_1d_2}{4\pi\epsilon_0}\sqrt{\frac{m^3\omega}{\hbar^5}}$ and $V_{SR}(\b{r}_1,\b{r}_2)$ is a potential of short-range interactions between molecules. For simplicity we assume that $V_{SR}(\b{r}_1,\b{r}_2)=V_{SR}(|\b{r}_1-\b{r}_2|)$

\begin{figure*}[ht]
\begin{center}
\includegraphics[width=0.65\textwidth]{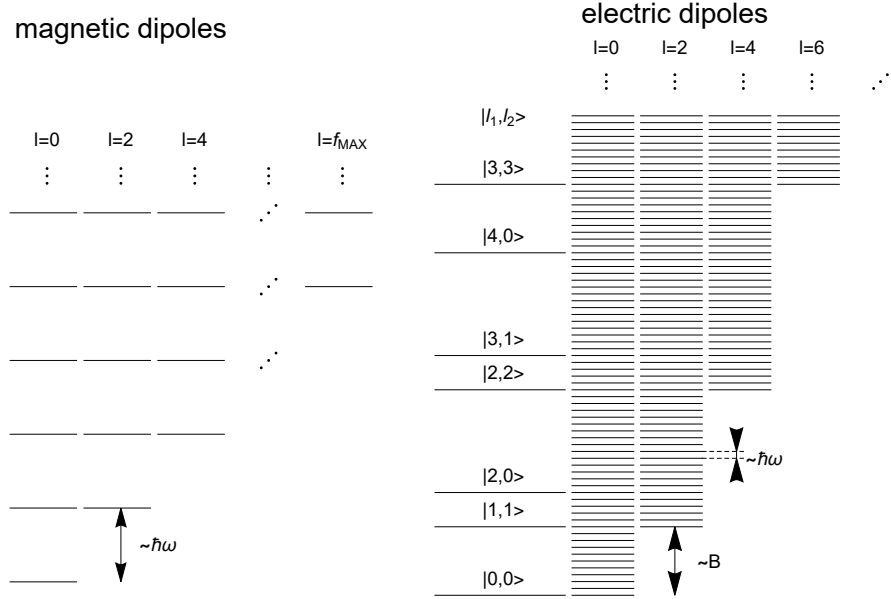}
\caption{Diagrams of energy levels for subspace which contains ground states of the whole system ($j=0$, $l=0$) for magnetic dipole and for bosonic electric dipoles (with no dipole-dipole interaction). For magnetic dipoles the energy of the first excited state with orbital angular momentum $l$ is very close to to the lowest state with orbital angular momentum $l+2$ (the only difference is caused by a short range interaction and it is much smaller than $\hbar\omega$). The $l$ is limited by maximal possible total spin of the system $f_{max}$. For electric dipoles, the oscillator energy $\hbar\omega$ is much smaller than the difference between different rotational states of molecule $\sim B$. What is more, $l$ is not limited here, as $l_1$ and $l_2$ may be arbitrarily large.}

\label{fig:dia}
\end{center}
\end{figure*}

\begin{figure*}[ht]
\begin{center}
\includegraphics[width=1\textwidth]{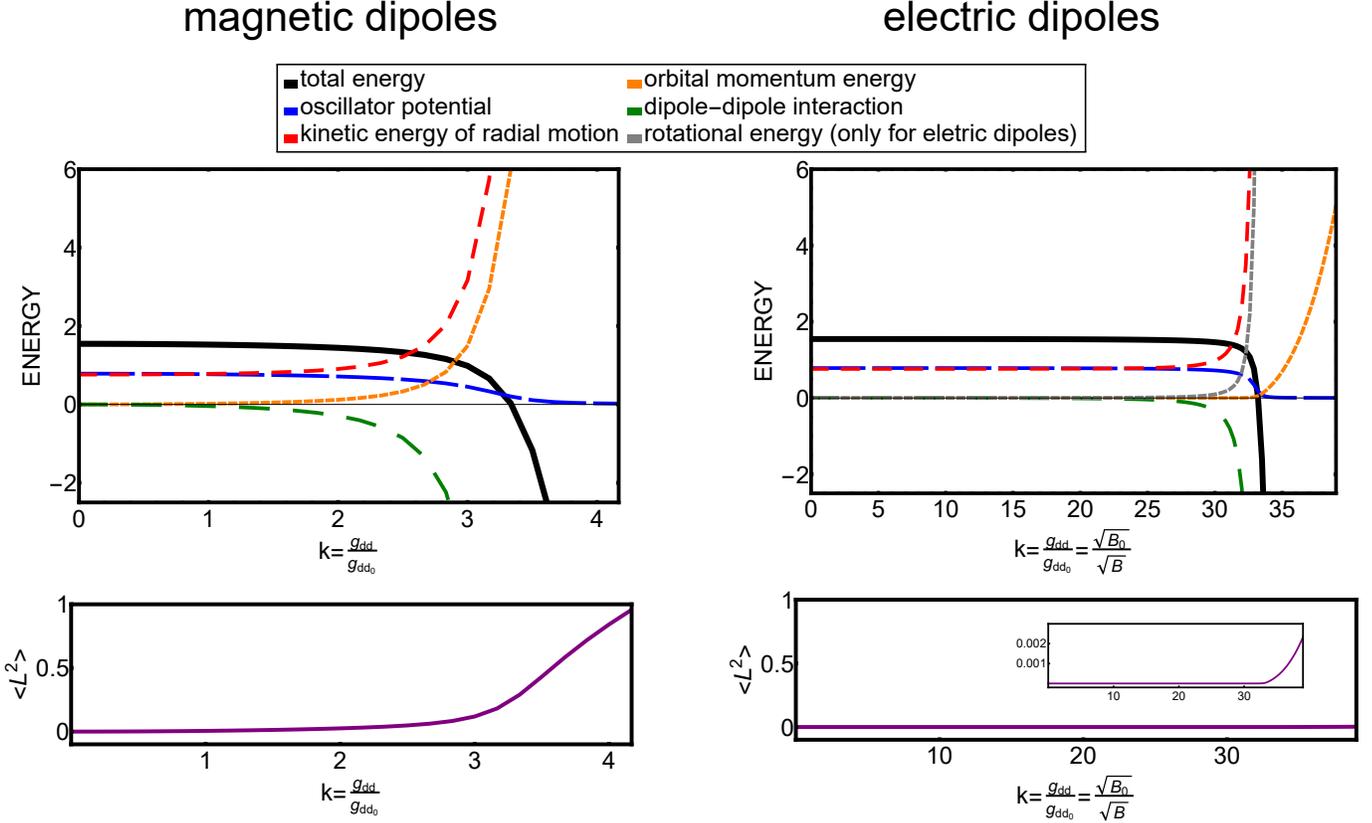}
\caption{Different parts of total energy $vs.$ $k$ coefficient for the ground states of the system of two dysprosium-like magnetic dipoles or two HCl-like electric dipoles. In both cases growing $k$ brings constituents closer, which implies increasing their kinetic energy.
In the magnetic case non negligible orbital angular momentum is generated, while in electric case this effect is infinitesimal.
}
\label{fig:em}
\end{center}
\end{figure*}

\begin{figure*}[ht]
\begin{center}
\includegraphics[width=1\textwidth]{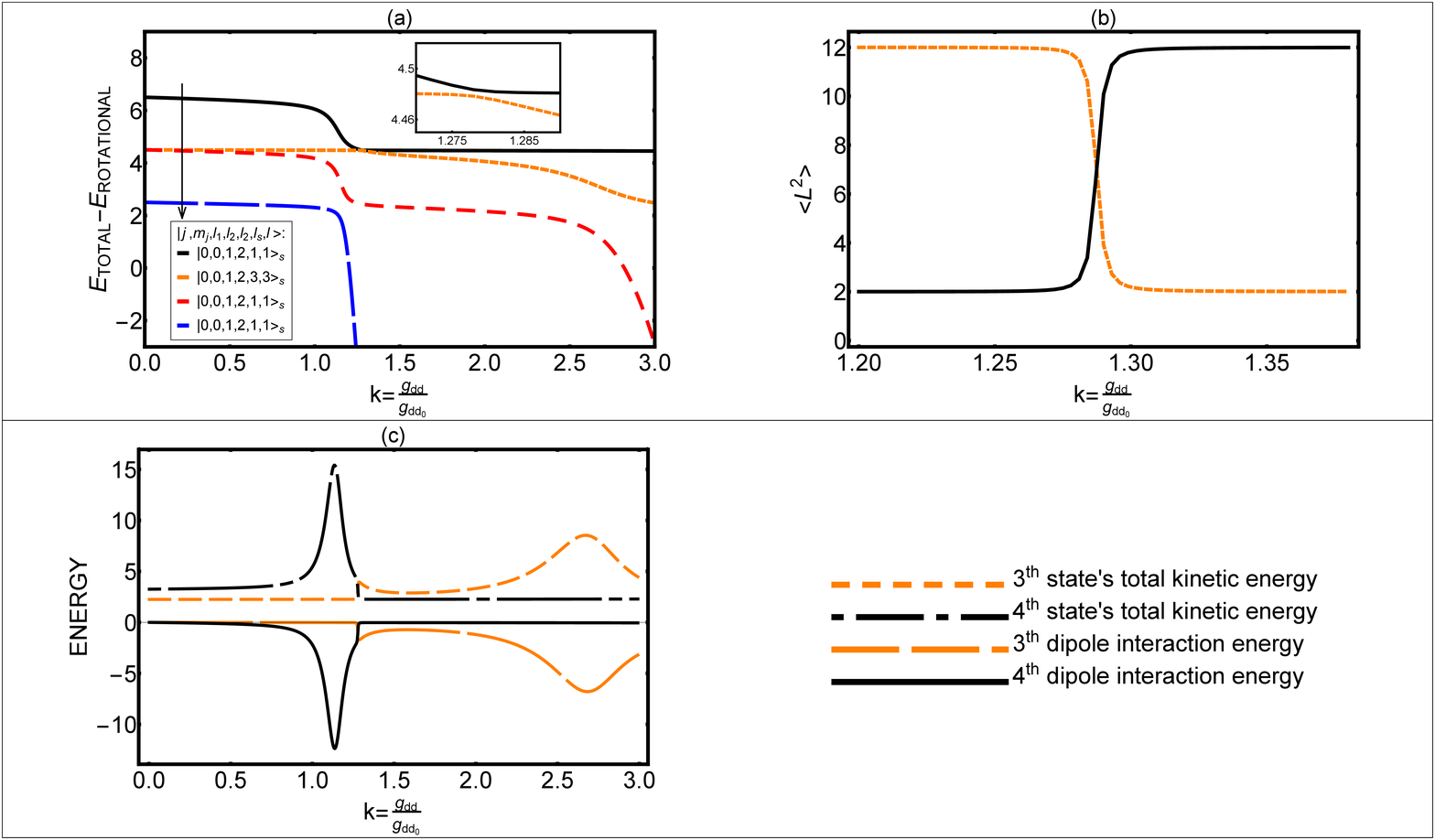}
\caption{The specific subspace of eigenstates is under consideration ($j=0$, $E_{rot}\approx 8B$, odd $l$). In plot (a) the energy $vs.$ $k$ is presented (as the $E_{rot}$ is fixed in this subspace, only other types of energy are presented). Characteristic anticrossing occurs, and the transfer of orbital angular momentum connected with it is presented at plot (b). In plot (c) dependence on $k$ for two parts of energy - total kinetic energy of relative motion (both radial and orbital) and dipole interaction energy - is plotted for two eigenstates.}
\label{fig:sub}
\end{center}
\end{figure*}

\section{Hamiltonian of relative motion}It is conventional to introduce vectors $\b{R}=\frac{1}{\sqrt{2}}(\b{r}_1+\b{r}_2)$ and $\b{r}=\frac{1}{\sqrt{2}}(\b{r}_1-\b{r}_2)$ and rewrite the Hamitlonian as:
\begin{equation}
H=H_{CM}+H_{rel,rot}
\end{equation}
where
\begin{align}
\begin{split}
H_{CM}&=-\frac{1}{2}\Delta_R+\frac{1}{2}R^2\\
H_{rel,rot}&=-\frac{1}{2}\Delta_r+\frac{1}{2}r^2+ B \b{L}_1^2+ B \b{L}_2^2\\
&+\frac{g_{dd}}{r^3}\big(\b{e}_1\cdot\b{e}_2-3(\b{e}_1\cdot\b{n})(\b{e}_2\cdot\b{n})\big)+V_{SR}(r)
\label{ham}
\end{split}
\end{align}
Here $g_{dd}=\dfrac{\tilde{g}_{dd}}{2^{3/2}}$. $H_{CM}$ is simply the harmonic oscillator Hamiltonian; now we will be interested in $H_{rel,rot}$. Note, that the part connected with the relative motion and the part connected with rotational states of molecules can not be easily separated because dipole-dipole interaction energy depends on both of them.\par
Angular momentum part of any eigenstate can be written down in a basis $|l,m_l,l_1,m_{1},l_2,m_{2}\rangle$, where $l,m_l$ are connected with relative motion and $l_1,m_{1},l_2,m_{2}$ describe rotational states of the molecules. Notice, that as the whole system is spherically symmetric, the total angular momentum is conserved:
\begin{equation}
[\b{J},H_{rel,rot}]=0
\end{equation}
(here $\b{J}=\b{L}+\b{L}_1+\b{L}_2$ and $\b{L}$ is the orbital angular momentum of relative motion). Therefore for our problem a basis $|j,m_j,l_1,l_2,l_s,l\rangle$ is a more appropriate choice (here $j,m_j$ are quantum numbers connected with $\b{J}$ operator and $l_s,m_{s}$ are connected with $\b{L}_s=\b{L}_1+\b{L}_2$). Of course the whole eigenfunction must be symmetric (or antisymmetric) with respect to exchange of particles. For states with $l_1\neq l_2$ we introduce index $p=s/a$:

\begin{multline}
|j,m_j,l_1,l_2,l_s,l\rangle_{s}\\=\frac{1}{\sqrt{2}}\big(|j,m_j,l_1,l_2,l_s,l\rangle+(-1)^l|j,m_j,l_2,l_1,l_s,l\rangle\big)
\end{multline}
or respectively:
\begin{multline}
|j,m_j,l_1,l_2,l_s,l\rangle_{a}\\=\frac{1}{\sqrt{2}}\big(|j,m_j,l_1,l_2,l_s,l\rangle+(-1)^{l+1}|j,m_j,l_2,l_1,l_s,l\rangle\big)
\end{multline}
Here, we have used the fact that the state of relative motion with even $l$ is symmetric (and the one with odd $l$ is antisymmetric). For states with $l_1=l_2$ the parity is determined by the sum $l+l_s$ (symmetric for even values or antisymmetric for odd).\par
As $j$, $m_j$ and parity $p=s/a$ are fixed (for given subspace of eigenstates), there remains four quantum numbers connected with the angular momenta ($l_1,l_2,l_s,l$). For such a subspace any eigenstate can be written as:
\begin{equation}
\Psi^{jm_jp}_n=\sum_{l_1l_2l_sl}\phi^{jm_jp}_{l_1l_2l_sln}(r)|jm_jl_1l_2l_sl\rangle_p
\label{sum}
\end{equation}
It is useful to introduce $\chi^{jm_jp}_{l_1l_2l_sln}(r)=r\phi^{jm_jp}_{l_1l_2l_sln}(r)$; then $\chi^{jm_jp}_{l_1l_2l_sln}(r)$ satisfy the set of equations:
\begin{multline}
-\frac{1}{2}\frac{d^2}{dr^2}\chi^{jm_jp}_{l_1l_2l_sln}(r)+\frac{1}{2}r^2\chi^{jm_jp}_{l_1l_2l_sln}(r)+\frac{l(l+1)}{2r^2}\chi^{jm_jp}_{l_1l_2l_sln}(r)\\+ B \big(l_1(l_1+1)+l_2(l_2+1)\big)\chi^{jm_jp}_{l_1l_2l_sln}(r)+
\\+\frac{g_{dd}}{r^3}\sum\limits_{l'_1l'_2l'_sl'}\alpha_{l_1l_2l_sl}^{l'_1l'_2l'_sl'jm_jp}\chi^{jm_jp}_{l'_1l'_2l'_sl'n}(r)+\\+V_{SR}(r)\chi^{jm_jp}_{l_1l_2l_sln}(r)=E^{jm_jp}_n\chi^{jm_jp}_{l_1l_2l_sln}(r)
\label{eq}
\end{multline}
where $\alpha_{l_1l_2l_sl}^{l'_1l'_2l'_sl'jm_jp}$ coefficients are the matrix elements:
\begin{multline}
\alpha_{l_1l_2l_sl}^{l'_1l'_2l'_sl'jm_jp}=\\
=\langle jm_jl_1l_2l_sl|_p[\b{e}_1\cdot\b{e}_2-3(\b{e}_1\cdot\b{n})(\b{e}_2\cdot\b{n})]|jm_jl'_1l'_2l'_sl'\rangle_p
\label{alf}
\end{multline}
From (\ref{alf}) one can see that
\begin{equation}
\alpha_{l_1l_2l_sl}^{l'_1l'_2l'_sl'jm_jp}\neq 0\Rightarrow (l-l')=\pm2\vee0
\end{equation}
so also the parity of $l$ is a constant of motion.

There are some restrictions for values of $l_1,l_2,l_s,l$:\\ $|l_1-l_2|\leq l_s\leq l_1+l_2$ and $|l_s-l|\leq j\leq l_s+l$, but still there is an infinite number of combinations $(l_1,l_2,l_s,l)$, so the sum (\ref{sum}) has infinitely many terms. The number of states which must be considered in the numerical calculation depends on the exact values of  $g_{dd}$ and $ B $ constants.

\section{Comparison to magnetic dipoles} The Hamiltonian of two magnetic dipolar atoms in a harmonic trap is very similar to the one of electric dipoles (it was analysed in some detail in \cite{my}):
\begin{align}
\begin{split}
H_{rel}^{mag}&=-\frac{1}{2}\Delta_r+\frac{1}{2}r^2+V_{SR}(r)\\
&+\frac{g_{dd}}{r^3}\big(\b{F}_1\cdot\b{F}_2-3(\b{F}_1\cdot\b{n})(\b{F}_2\cdot\b{n})\big)
\label{ham}
\end{split}
\end{align}
were $\b{F}_1,\b{F}_2$ are spin (total internal angular momentum) operators. There are two significant differences: firstly there is no term responsible for rotational energy, because the orientation of spin is not connected with atom's moment of interia. Secondly, spins of both single particles $f_1,f_2$ are fixed, so there remains only four quantum numbers related to the angular momenta $|l,m_l,m_{1},m_{2}\rangle$. Let us introduce the operator $\b{F}=\b{F}_1+\b{F}_2$ (quantum numbers $f,m_f$) and write down the sum (analogous to (\ref{sum})):
\begin{equation}
\Psi^{jm_j}_n=\sum_{fl}\phi^{jm_j}_{fln}(r)|jm_jfl\rangle
\end{equation}
(here $f$ and $l$ determine the parity of state, so no additional index $p$ is needed). In opposite to (\ref{sum}), this sum is finite, as $f\leq f_1+f_2$, $|j-f|\leq l\leq j+f$.\par
In both magnetic and electric cases the total angular momentum $j,m_j$, parity of the whole state $p$ and parity of the orbital angular momentum number $l$ are constants of motion. Due to this fact it is useful to use basis $|j,m_j,l_1,l_2,l_s,l\rangle_p$ (for electric dipoles) or $|j,m_j,f,l\rangle$ (magnetic). Before we start looking for eigenstates it is worth taking a look at this subspaces with no dipole-dipole interaction. In Fig. \ref{fig:dia} there are schemes of subspaces which contain the ground states of the whole Hamiltonian for bosons ($j=m_j=0$, symmetrical states, even $l$). For magnetic dipoles the value of the energy interval between states with different $l$ is $\sim\hbar\omega$ (it is slightly perturbed by the short range interaction). The value of $l$ is limited by maximum possible value of $f$ (with a given parity). For electric dipoles the value of energy interval is much bigger and it is connected with rotational energy of particles (as whole $j$ is fixed, $l\neq 0$ implies $l_1,l_2\neq 0$). Note, that the rotational state $|l_1,l_2\rangle$ may appear in a given column only if $|l_1-l_2|\leq l \leq l_1+l_2$. In the electric case $l$ is not limited. The other subspaces (with different values of $j,m_j$ and parities) have a very similar structure.

\section{Ground state}Now we want to investigate the eigenstates of the whole Hamiltonian. We start with the ground state of the system of two bosonic HCl-like molecules in the harmonic trap with a frequency $2.8\cdot2\pi\quad kHz$. For such a situation values of constants are $g_{dd}=0.27$ and $B=1.15\cdot10^8$. For numerical calculations we use hard-core model of short-range potential $V_{SR}(r)$:
\begin{equation}
    V_{SR}(r):= \begin{cases}
               +\infty               & for\quad r<b\\
               0               & for\quad r>b\\
           \end{cases}
\end{equation}
and we assume that $b\approx a_0$, where $a_0$ is a typical value of the scattering length for dipolar molecules (approximately equal to one hundred Bohr radii $100r_0$\cite{sr}).
Of course it is oversimplified. However, in our case the exact form of $V_{SR}(r)$ is not crucial. We want to focus our attention on the dipole-dipole interaction; what is more, this approach lets us solve the system of equations (\ref{eq}) using the shooting method. In harmonic oscillator units $b=0.04$.\par

The ground state of the whole system is the one with dominating angular momentum part $|j,m_j,l_1,l_2,l_s,l\rangle=|0,0,0,0,0,0\rangle$. As it is far away in energy from any other rotational state, in this case the dipole-dipole interaction may be treated as a perturbation. As a result, the additional effective potential
\begin{equation}
V_{eff}(r)=-\frac{g_{dd}^2(\alpha^{112200p}_{0000})^2}{6B}\frac{1}{r^6}
\end{equation}
occurs in the equation for radial function $\chi^{00p}_{00000}$ (which is a correction to the Van der Waals interaction potential; this effect is well known, see for instance \cite{ll}) and the radial function connected with $l=2$ is approximately equal:
\begin{equation}
\chi^{00p}_{11220}(r)\approx-\frac{g_{dd}\alpha^{112200p}_{0000}}{6B}\frac{1}{r^3}\chi^{00p}_{00000}(r)
\end{equation}
All functions connected with higher $l$ are negligible here.

We are interested in investigating how the ground state's quantities depend on the strength of the dipole-dipole interaction. This strength may be effectively changed by manipulating trap frequency. Also $B$ value depends on $\omega$, but with different power. By using Feshbach resonances \cite{f1,f2,f3,f4} we are able to keep $b$ constant. Finally, we can introduce coefficient $k=\frac{g_{dd}}{g_{dd_0}}=\frac{\sqrt{B_0}}{\sqrt{B}}$ and investigate ground state dependence on $k$.

In Fig. \ref{fig:em} we see, how different types of energy are changing with $k$ for the system of two dysprosium-like magnetic dipoles or two HCl-like electric dipoles. When dipole-dipole interaction becomes strong enough, molecules attract each other getting closer. It is of course connected with a rising kinetic energy (due to the Heisenberg uncertainty principle), decreasing of the total energy and  increasing mean value of $\langle L^2\rangle$.
The $\langle L^2\rangle$ generated for electric dipoles is very small (in comparison with the case of magnetic dipoles), due to the fact, that the state with a non-zero $\langle L^2\rangle$ has huge rotational energy here.\par

We see that for$k\geq33$ the system cannot be treated as two individual particles any more (rigid rotators approximation is not valid here) - it corresponds to the fact, that in nature particles with big values of the electric dipole moment and the moment of interia form a solid, not a gas.

When we want to consider fermionic case, the reasoning is almost identical (there for the ground state $|j,m_j,l_1,l_2,l_s,l\rangle=|0,0,0,0,0,1\rangle$, but this is the only significant difference).

\section{Subspaces of eigenstates} We are interested in the regime in which the system can be treated as two individual particles. In this regime rotational energy is much bigger than the energy of dipole-dipole interaction (even if the distance between particles is very close to $b$ we have $B> 10^4\dfrac{g_{dd}}{b^3}$). 
From Fig. \ref{fig:dia} one can see that for any state with given rotational energy $E_{rot}= B \cdot\big(l_1(l_1+1)+l_2(l_2+1)\big)$ (i.e. $l_1=1,l_2=1$) there exist states with with smaller $l_1,l_2$ numbers (i.e. $l_1=0,l_2=0$), which are very close in total energy to the first ones. However, these are highly excited harmonic oscillator states, so the overlap of radial functions of states with $l_1=0,l_2=0$ with radial functions of the lowest states with $l_1=1,l_2=1$ will be negligible. Therefore it is reasonable to assume that the eigenstates are very close to the states with well-defined rotational energy $E_{rot}= B \cdot\big(l_1(l_1+1)+l_2(l_2+1)\big)$ and to treat dipole-dipole interaction as a small perturbation. It's worth pointing out that $E_{rot}$ does not determine values of $l_1,l_2$ - for example $l_1=0,l_2=3$ and $l_1=2,l_2=2$ give us the same $E_{rot}=12 B $.\par
Let us consider the subspace with given $j,m_j,p$ and the energy close to some $E_{rot}$. There are two possible situations - the first one: there are no two angular momentum states $|j,m_j,l_1,l_2,l_s,l\rangle_p$, $|j,m_j,l'_1,l'_2,l'_s,l'\rangle_p$ for which $|l_1-l'_1|=1$ and $|l_2-l'_2|=1$ in this subspace (the example of such a subspace is the one mentioned in the previous paragraph). Then the states of this subspaces do not interact with each other; they can be only weakly coupled to the states from other subspaces. The example of that type of subspace is the one which contains the ground state.
In this case the effects of dipole-dipole interaction reduces here to the correction to ordinary Van der Waals forces.\par
The second situation is when there exist two angular momentum states $|j,m_j,l_1,l_2,l_s,l\rangle_p$, $|j,m_j,l'_1,l'_2,l'_s,l'\rangle_p$ for which $|l_1-l'_1|=1$ and $|l_2-l'_2|=1$ (for example $E_{rot}=9 B $, $l_1=1,l_2=2,l'_1=2,l'_2=1$). In this case we can observe strong electric dipole-dipole interaction. If we neglect (very weak) coupling to states from different subspaces, the sum in equation (\ref{sum}) becomes finite and the whole problem is reduced to a finite number of equations for radial functions (eq. \ref{eq}).\par
As long as we stay in subspace with fixed $l_1(l_1+1)+l_2(l_2+1)$, the rotational energy may be ignored. Within this subspace molecule has some freedom in orientation. As this subspace contains only finite rotational states of molecules, effectively their dipolar momenta are quantizied (in some sense analogously to magnetic dipoles). Indeed, subspaces with fixed $l_1(l_1+1)+l_2(l_2+1)$ turn out to by very similar to the subspaces of eigenstates of magnetic dipoles (for which spins $f_1$, $f_2$ are fixed by their nature). In Fig. \ref{fig:sub} we analyze subspace of angular momentum states $|0,0,1,2,3,3\rangle_s,|0,0,1,2,1,1\rangle_s$. In plot (a) the energies of four lowest states in this subspace are presented. For small $k$ they are very close to the ones with well defined $l_1,l_2,l_s,l$ quantum numbers. For $k\approx0.2$ the state with $l=1$ is slightly lower in energy than corresponding state with $l=3$, due to dipole-dipole interaction - this effect is stronger here than the energy shifts generated by short-range interaction (which increase $l=1$-state's energy mainly). The plot is very similar to the one from Ref. \cite{my}. For $k\approx 1.3$ very narrow anticrossing between the $3^{th}$ state and the $4^{th}$ state occurs. It is connected with strong orbital angular momentum transfer, which is presented in plot (b). While the presence anticrossing in a situation, when lines of eigenenergies are getting closer to each other is well known effect \cite{anty1,anty2}, the interesting question is: what physical effect makes the energy of the state with dominating $|0,0,1,2,1,1\rangle_s$ angular momentum part decreasing, what leads to this anticrossing. To answer this question, in the plot (c) we compare the energies - total kinetic energy of the relative motion (which includes both: radial and orbital energies) and the energy of dipole-dipole interaction - for two states which participate in the anticrossing (the oscillator energy does not depend strongly on $k$ and it is omitted here for clarity) . As one can see, decreasing of energy is connected with some kind of resonance, where particles get closer to each other, what causes increasing kinetic energy and strengthens attractive dipole interaction. Analogous, but not so rapidly, effects occurs also for higher $k$ ($\approx 2.7$).

\section{Conclusions}We have analyzed the motion of two molecules with permanent electric dipole moment moving in the spherical harmonic trap. The Heisenberg uncertainty principle manifests itself here: to orient a dipole in a given direction a great orbital energy is needed. Consequently, the ability of choosing direction is highly restricted by the energy of rotational state, which is really huge compared to the energy of electric interactions between molecules. As a result, the electric dipole interaction (which classically is much stronger than magnetic) has significant impact on the behaviour of the system only in some very specific cases. Especially, ground state of the system of two identical electric dipolar molecules under typical conditions for gaseous phase almost does not feel the dipolar interaction. Extension of $k$ leads to a collapse of the system. For particles which naturally form a gas the whole subspace of eigenstates may be separated into parts with well defined rotational energy; these parts are very similar to the spaces of eigenstate for magnetic dipoles and there exist subspaces for which strong transfer of $\langle L^2\rangle$ may be observed. However, it will be rather hard to be observed in experiment, as it require preparation of the system in very special excited state.

\acknowledgments
The authors are pleased to thank R. O\l{}dziejewski for contribution to the numerical method used in this paper. This work was supported by (Polish) National Science Center Grant 2015/19/B/ST2/02820.

\end{document}